\documentclass[floatfix,twocolumn,prb,amsmath,amssymb]{revtex4}

\usepackage{graphicx}
\usepackage{dcolumn}
\usepackage{bm}
\usepackage[section]{placeins}

\def\R{{\mathbb R}}

\def\d{{\rm d}}

\begin{document}

\title{Nonadiabatic nuclear dynamics of the ammonia cation studied by surface hopping classical trajectory calculations}

\author{Andrey K. Belyaev}
\email{belyaev@herzen.spb.ru}
\affiliation{Department of Theoretical Physics, Herzen University, St. Petersburg 191186, Russia}

\author{Wolfgang Domcke}
\email{wolfgang.domcke@ch.tum.de}
\affiliation{Department Chemie, Technische Universit\"at M\"unchen, D-85747 Garching, Germany}

\author{Caroline Lasser}
\email{classer@ma.tum.de}
\affiliation{Zentrum Mathematik, Technische Universit\"at M\"unchen, D-85747 Garching, Germany}

\author{Giulio Trigila}
\email{trigila@ma.tum.de}
\affiliation{Zentrum Mathematik, Technische Universit\"at M\"unchen, D-85747 Garching, Germany}

\date{\today}

\begin{abstract}
The Landau--Zener (LZ) type classical-trajectory surface-hopping algorithm is applied to the nonadiabatic nuclear dynamics of the ammonia cation after photoionization of the ground-state neutral molecule to the excited states of the cation.
The algorithm employs the recently proposed formula for nonadiabatic LZ transition probabilities derived from the adiabatic potential energy surfaces.
The evolution of the populations of the ground state and the two lowest excited adiabatic states is calculated up to 200 fs.
The results agree well with quantum simulations available for the first 100 fs based on the same potential energy surfaces.
Four different time scales are detected for the nuclear dynamics:
Ultrafast Jahn--Teller dynamics between the excited states on a 5 fs time scale;
fast transitions between the excited state and the ground state within a time scale of 20 fs;
relatively slow partial conversion of a first-excited-state population to the ground state within a time scale of 100 fs;
and nearly constant populations  after roughly 120 fs due to a dynamical equilibrium between all three states.
The latter provides a possible explanation of the experimental evidence that ammonia cation is nonfluorescent.
\end{abstract}

\maketitle

\section{Introduction}

Many important physical processes and chemical reactions involve nonadiabatic transitions between adiabatic electronic states, often mediated by conical intersections \cite{DYK:2004, Bersuker:2006, DYK:2011, DomckeYakony:2012}. An interesting example is the photoionization of ammonia (NH$_3$) and the nonadiabatic nuclear dynamics of the ammonia cation (NH$_3^+$)  which for decades have been of great interest from the experimental as well the theoretical point of view, see, e.g.,  Refs.~
\cite{Haller-etc:1980cpl, Edvardsson:1999jpb, Woywod-etc:2003jcp, Viel-etc:2006jcp, Viel-etc:2008cp, Webb-etc:2009jpca} and references therein.

Nonadiabatic electronic transitions are quantum phenomena and, in principle, should be studied by means of quantum mechanical methods.  
Nonadiabatic effects can be investigated in detail for small systems with quantum mechanical methods. 
However, quantum calculations are costly or may be even impossible for the nonadiabatic nuclear dynamics of somewhat larger molecules. 
For these cases, more approximate classical or semiclassical methods offer an important alternative because of their lower computational cost and the physical insight they provide into the dynamics of a reaction;  
see the special issue dedicated to the nonadiabatic nuclear dynamics headed by the Perspective \cite{Tully:2012}. 
Of particular interest are mixed quantum--classical approaches which treat the electronic motion quantum mechanically and the nuclear motion classically. 

Among the many quasi-classical methods for treating the nonadiabatic nuclear dynamics, e.g., 
the semiclassical initial-value representation (IVR) \cite{Miller:1970, Marcus:1972, KreekMarcus:1972, Miller:2001}, 
the Ehrenfest dynamics method \cite{McLachlan:1964, MeyerMiller:1979, Micha:1983, Kirson-etc:1984, Sawada-etc:1985}, 
the frozen Gaussian wave-packet method \cite{Heller:1991}, 
the multiple-spawning wave-packet method \cite{Ben-NunMartinez:1998, Martinez-etc:2000, Ben-NunMartinez:2002, Martinez-etc:2012}, 
to mention a few, 
the classical trajectory surface-hopping method with its many variants \cite{BjerreNikitin:1967, TullyPreston:1971, MillerGeorge:1972, StineMuckerman:1976, Kuntz-etc:1979, BlaisTruhlar:1983, Tully:1990, HammesSchiffer-Tully:1994, MuellerStock:1997, Voronin-etc:1998, Thiel-etc:2008, FermanianLasser:2008, LasserSwart:2008, Belyaev-etc:2014jcp} 
is one of the most widely used mixed quantum-classical computational methods. 

The key feature distinguishing different surface-hopping methods is the way of calculating nonadiabatic transition probabilities. 
The original fewest-switches approach \cite{Tully:1990} solves the time-dependent Schr\"odinger equation along classical trajectories in combination with the probabilistic fewest-switches algorithm at each integration time step to make a decision whether to switch the electronic state or not. 
This allows one to treat the nonadiabatic nuclear dynamics without determining nonadiabatic regions beforehand, but specify them along each treated classical trajectory. 
A widely used alternative is to use a nonadiabatic model, typically, the Landau-Zener (LZ) model, for the calculation of nonadiabatic transition probabilities, see, e.g., Refs.~\cite{BjerreNikitin:1967, TullyPreston:1971, Kuntz-etc:1979, Voronin-etc:1998, Thiel-etc:2008, FermanianLasser:2008, BelyaevLebedev:2011, Belyaev-etc:2014jcp}. 
In practical applications, the challenge of using the Landau-Zener model is two-fold: (i) to find nonadiabatic regions where a hopping should take place, and (ii) to calculate a LZ parameter and a LZ nonadiabatic transition probability in each particular nonadiabatic region and for each particular classical trajectory. 
Usually, the former requires beforehand analysis, while the latter needs a diabatization procedure, which is not uniquely defined, especially if the nonadiabatic coupling element is unknown. 
These problems can often make the practical application of the LZ model difficult in multi-dimensional applications. 

Both problems can be solved by means of the recently derived adiabatic-potential-based formula \cite{BelyaevLebedev:2011, Belyaev-etc:2014jcp} within the LZ model. The center of a nonadiabatic region is determined by a local minimum of the separation of adiabatic potential energies along a classical trajectory, and the LZ nonadiabatic transition probability in this region is calculated based on the time-dependent adiabatic energy gap, see below. 
No diabatization procedure is required. 
The adiabatic-potential-based formula \cite{BelyaevLebedev:2011} derived within the LZ model is somehow similar, but not identical, to a nonadiabatic transition probability formula obtained by an analytic continuation by \citet{MillerGeorge:1972} . 
This LZ surface hopping algorithm has very recently been tested for the two-dimensional two-mode model of a conical intersection \cite{Belyaev-etc:2014jcp} and compared with other numerical algorithms and quantum dynamics calculations. 
In the present paper, this algorithm is applied to the nonadiabatic nuclear dynamics of the ammonia cation based on an accurate six-dimensional three-sheeted adiabatic potential energy surface (PES) \cite{Viel-etc:2006jcp} which involves conical intersections of both Jahn-Teller and pseudo-Jahn-Teller types.

Photoelectron spectra of ammonia corresponding to the electronic ground state, as well as the two-fold degenerate excited state of the ammonia cation have been measured, see \cite{Edvardsson:1999jpb, Webb-etc:2009jpca} and references therein for experimental results. 
One of the interesting  experimental findings is that NH$_3^+(\tilde A~^2E - \tilde X~^2A_1)$ fluorescence was not observed \cite{Dujardin:1985}, supporting the idea of fast nonradiative decay processes due to a conical intersection \cite{Krier:1985}. 
A PES based  linear vibronic coupling model for the $\tilde X~^2A_1$ state and the first excited $\tilde A~^2E$ state \cite{Woywod-etc:2003jcp} was used to perform the first quantum wave packet calculation. 
In 2006, an analytical diabatic six-dimensional three-sheeted PES for the ground and two lowest-lying excited states of the ammonia cation was developed based on accurate ab initio multireference configuration interaction calculations.
This model includes higher-order coupling terms both for the Jahn-Teller and pseudo-Jahn-Teller matrix elements. 
Ultimately, six-dimensional wave packet dynamics calculations were performed employing the multiconfigurational time-dependent Hartree (MCTDH) method were performed up to 100~fs \cite{Viel-etc:2006jcp}. 
These dynamics calculations explained most experimental evidences, in particular the complex vibronic structure of the photoelectron spectra.
Nevertheless, there still exists an open question. 
It was found that about 30\% of the $\tilde A$ state population did not decay to the $\tilde X$ state of the cation within 100~fs after photoionization.

More accurate quantum dynamics calculations based on the same PESs did not lower this fraction. On the contrary, the $\tilde A$-state population was found to be of 40\% after 100~fs \cite{Viel-etc:2008cp}. 

Therefore, a first principles theoretical study of the nonadiabatic nuclear dynamics of the ammonia cation is still of great interest. 
The surface hopping algorithm based on the adiabatic-potential-based formula \cite{BelyaevLebedev:2011, Belyaev-etc:2014jcp} is attractive for this purpose, since is well suited for a nonadiabatic dynamics study of a multi-dimensional system which exhibits conical intersections between several electronic states. 
The given diabatic representation of the six-dimensional three-sheeted ammonia-cation PES \cite{Viel-etc:2006jcp} allows one to calculate three corresponding adiabatic PESs, which are needed for the classical trajectory propagation and the adiabatic-potential-based formula for nonadiabatic transition probabilities, with minimal computational cost.
Moreover the classical trajectory surface hopping results can be compared with converged quantum dynamics calculations \cite{Viel-etc:2006jcp, Viel-etc:2008cp} on the same PESs. 
%
%
This provides a stringent test of the accuracy of the classical trajectory surface hopping approach for a nontrivial polyatomic system. Due to the lower computational cost, the surface hopping calculations can cover a longer time scale and provide physical insight into the
nonadiabatic dynamics beyond the first hundred femtoseconds.

\section{The model}

We describe the nuclear positions of NH$_3^+$ by a Cartesian coordinate vector  
$q\in\R^{d}$, $d=12$, and group the coordinates of the four atoms as 
$$
q = (q_1,q_2,q_3,q_4)\in\R^3\times\cdots\times \R^3.
$$
In Cartesian coordinates, the kinetic energy operator has the simple form
\begin{equation}\label{eq:kin}
T = \sum_{j=1}^4  -\frac{\hbar^2}{2m_j} \Delta_{q_j},
\end{equation}
where $m_1,\ldots,m_4$ denote the masses of the four atoms. 

\subsection{The potential energy matrix}
For the potential energy operator  we use the $3\times 3$ diabatic potential energy matrix 
$$
V = V^{\rm diag} + V^{\rm coup}
$$
developed in Ref.~\cite{Viel-etc:2006jcp}. This real symmetric matrix is expressed in six-dimensional symmetry adapted internal coordinates $S=(S_1,\ldots,S_6)$: the symmetric stretch~$S_1$, the umbrella coordinate~$S_2$, the asymmetric stretching and bending coordinates $S_3,S_4$ and $S_5,S_6$, respectively.
The diabatic coupling matrix combines a Jahn--Teller with a 
pseudo Jahn--Teller matrix,
$$
V^{\rm coup} =  \begin{pmatrix}0 & W^{\rm PJT} &- Z^{\rm PJT}\\ W^{\rm PJT} & W^{\rm JT} & Z^{\rm JT}\\ - Z^{\rm PJT} & Z^{\rm JT} & -W^{\rm JT}\end{pmatrix}.
$$
The diabatic coupling matrix $V$ has three real eigenvalues 
$$
\lambda_1\le \lambda_2\le \lambda_3,
$$ 
the adiabatic ground state PES $\lambda_1$ and the excited state surfaces $\lambda_2$ and $\lambda_3$.
We denote the corresponding normalized eigenvectors by $\chi_{j}$, such that
$V \chi_j = \lambda_j \chi_j$ for $j=1,2,3$. 
The Landau--Zener surface hopping algorithm only requires the adiabatic PES and their energy gaps 
$$
Z_{12} = \lambda_2-\lambda_1,\qquad Z_{23} = \lambda_3-\lambda_2,
$$
but not the diabatic matrix $V$.  

\subsection{The initial wave function}

The initial three-level wave function $\psi_0$ for the solution of the time-dependent Schr\"odinger equation
$$
i\hbar \partial_t\psi_t = (T+V) \psi_t
$$
is the vertically excited, one-level ground state $\phi_{\rm neut}$ of neutral ammonia NH$_3$, which equally populates the excited states $\chi_{2}$ and $\chi_{3}$ of the cation NH$_3^+$.

The three-level wave function at time $t$ is given as
$$
\psi_t = \psi_t^{(1)}\chi_1 + \psi_t^{(2)}\chi_2 + \psi_t^{(3)}\chi_3
$$
with adiabatic one-level wave functions $\psi_t^{(j)}$, $j=1,2,3$, and the initial condition
\begin{equation}\label{eq:initial}
\psi_0^{(1)} = 0,\quad \psi_0^{(2)} = \psi_0^{(3)} = \phi_{\rm neut}/\sqrt{2}.
\end{equation}

\subsection{Approximating the initial wave function}

The one-level Schr\"odinger operator of neutral ammonia,
$T + V_{\rm neut}$, uses the potential function $V_{\rm neut}$ of Ref.~\cite{Viel-etc:2006jcp}, which is constructed with respect to the same six-dimensional symmetry 
adapted  coordinates as the diabatic matrix $V$. The neutral ground state $\phi_{\rm neut}$ is approximated by the ground state $\phi_{\rm harm}$ of the harmonic Schr\"odinger operator
\begin{equation}\label{eq:harm}
T + \tfrac12 (q-q_*)\cdot D(q-q_*),
\end{equation}
where $q_*$ is the equilibrium configuration of neutral ammonia and $D$ is the $12\times 12$ diagonal matrix defined by the diagonal components of the Hessian matrix of $V_{\rm neut}$ evaluated in $q_*$.

\subsection{The initial condition in phase space}

The Wigner functions of the adiabatic one-level functions $\psi^{(j)}_t$ are defined as
\begin{eqnarray*}
\lefteqn{W(\psi^{(j)}_t)(q,p) =}\\
&& (2\pi\hbar)^{-d} \int e^{i y\cdot p} \psi_t^{(j)}\!\left(q-y/2\right)  \psi_t^{(j)}\!\left(q+ y/2\right)^* \d y.  
\end{eqnarray*}
They map phase space points $(q,p)\in\R^{2d}$, $d=12$, to the real numbers. For the initial adiabatic wave functions of equation (\ref{eq:initial}), we have
$$
W(\psi^{(1)}_0)=0,\quad
W(\psi^{(2)}_0) = W(\psi^{(3)}_0) = W(\phi_{\rm neut})/2
$$
with the phase space Gaussian
\begin{eqnarray}
\lefteqn{W(\phi_{\rm neut})(q,p) =}\label{eq:wigner}\\
&& \frac{1}{(\pi\hbar)^{d}} \exp\!\left(-\left((q-q_*)\cdot D_m(q-q_*) +p\cdot  D_m^{-1} p\right)/\hbar\right)
\nonumber
\end{eqnarray}
and the the $12\times 12$ diagonal matrix
$$
D_m = \left({\rm diag}(m_1,m_1,m_1,\ldots,m_4,m_4,m_4) D\right)^{1/2},
$$
which is obtained by the appropriate mass scaling of the diagonal matrix $D$ defined by the harmonic approximation in equation (\ref{eq:harm}).

\section{Surface hopping trajectories}

The Landau--Zener surface hopping algorithm \cite{Belyaev-etc:2014jcp} is used to compute level populations as well as position 
expectations for the nonadiabatic dynamics of the ammonia cation in Cartesian coordinates, benefiting from the simple form of the kinetic energy operator in Eq.~(\ref{eq:kin}).

\subsection{Initial sampling}
Initially, the second and the third adiabatic levels are populated, and we choose phase space sampling points 
$$
(q_1^{(j)},p_1^{(j)}),\ldots,(q_{N_0}^{(j)},p_{N_0}^{(j)})\in\R^{2d}, \qquad j=2,3,
$$ 
$d=12$, from the initial Gaussian Wigner functions given 
in equation (\ref{eq:wigner}). Two sets of $2d$-dimensional Halton points, which deterministically 
approximate the uniform distribution on the unit cube $[0,1)^{2d}$, are mapped by the inverse of cumulative 
distribution functions of univariate normal distributions to approximate the multivariate Gaussian Wigner functions. 
The corresponding quasi-Monte Carlo estimate then provides
$$
\int A(q,p) W(\psi_0^{(j)})(q,p) d(q,p) \approx \frac{1}{2N_0}\sum_{k=1}^{N_0} A(q_k^{(j)},p_k^{(j)}) 
$$
with an error of the order $(\log N_0)^{2d}/N_0$, when integrating over a phase space function $A$ with respect 
to the adiabatic Wigner functions.

\subsection{Classical trajectories}
Let $(q,p)=(q_1,\ldots,q_4,p_1,\ldots,p_4)\in\R^{24}$ be a phase space point associated with the $j$th adiabatic level. 
We evolve the point according to the classical Hamiltonian system
$$
\dot q_k = \frac{1}{m_k}\, p_k,\qquad \dot p_k = -\partial_{q_k} \lambda_j(q)
$$
for all coordinates $k=1,\ldots,4$. The discretization uses a symplectic fourth order Runge--Kutta scheme.

\subsection{Landau--Zener transitions}
Whenever one of the eigenvalue gaps becomes minimal along an individual classical trajectory a nonadiabatic 
transition occurs. 

Let us consider a classical trajectory $t\mapsto (q(t),p(t))$ associated with the first
adiabatic surface corresponding to level $\chi_{1}$. Whenever the gap function $t\mapsto Z_{12}(q(t))$ attains a local minimum, a transition to 
the second adiabatic surface corresponding to level $\chi_{2}$ might be performed.
We denote such a critical point of time by $t_c$ and the corresponding phase space point by $(q_c,p_c)$. 
We evaluate the Landau--Zener probability\cite{BelyaevLebedev:2011}
$$
P_{\rm LZ} = \exp\!\left(-\frac{\pi}{2\hbar}\sqrt{\frac{Z_{12}(q_c)^3}{\frac{d^2}{dt^2} Z_{12}(q(t))\mid_{t=t_c}}}\right),
$$
and compare with a pseudorandom number $\xi$ generated from the uniform distribution on $[0,1]$. 
If $\xi\le P_{\rm LZ}$, then the trajectory hops on the  second adiabatic surface with rescaled momentum $p_{\rm rs}$ such 
that
$$
\frac12|p_c|^2 + \lambda_1(q_c) = \frac12|p_{\rm rs}|^2 + \lambda_2(q_c).
$$
If $\xi>P_{\rm LZ}$, then the classical trajectory remains on the first surface. 

Classical trajectories associated with the second adiabatic surface are treated analogously, however, monitoring both the gap 
functions $Z_{12}$ and $Z_{23}$. For classical trajectories running on the third surface only the gap function $Z_{23}$ is relevant.

\subsection{Evaluation of the observables}
At some time $t$, the surface hopping algorithm has produced three sets of phase space points
$$
(q_1^{(j)},p_1^{(j)}),\ldots,(q_{N_j(t)}^{(j)},p_{N_j(t)}^{(j)})\in\R^{2d}, \qquad j=1,2,3,
$$
such that the total number of points is constant over time, $N_1(t) + N_2(t) + N_3(t) = 2N_0$. The adiabatic level populations are approximated 
by counting the phase space points, that is, 
$$
\langle \psi_t^{(j)}\mid \psi_t^{(j)}\rangle  \approx  \frac{N_j(t)}{2N_0}.
$$
The position expectations at time $t$ are deduced from the arithmetic means on each adiabatic level, that is,
$$
\langle \psi_t^{(j)}\mid \hat q \mid \psi_t^{(j)} \rangle \approx  \frac{1}{N_j(t)} \sum_{k=1}^{N_j(t)} q_k^{(j)}.
$$

\section{Results}
We now report the results of the Landau--Zener type surface-hopping algorithm described above to the nonadiabatic nuclear dynamics of the ammonia cation after photoionization of ammonia  for a time interval of $200$ fs.

\subsection{Adiabatic population and coordinate evolution}
\begin{figure}[t]
    \begin{tabular}{l}
    \resizebox{70mm}{!}{\includegraphics{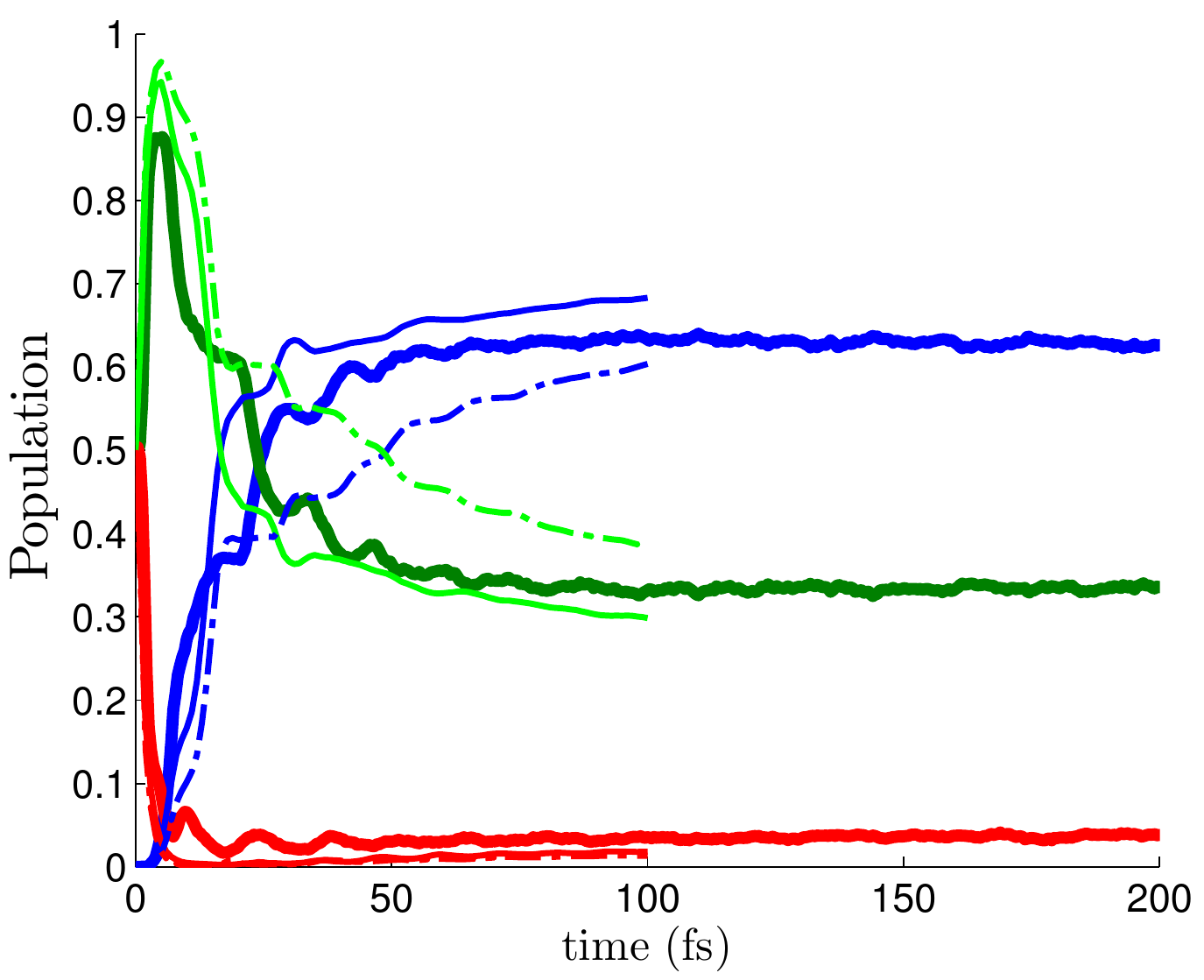}}\\
    \resizebox{70mm}{!}{\includegraphics{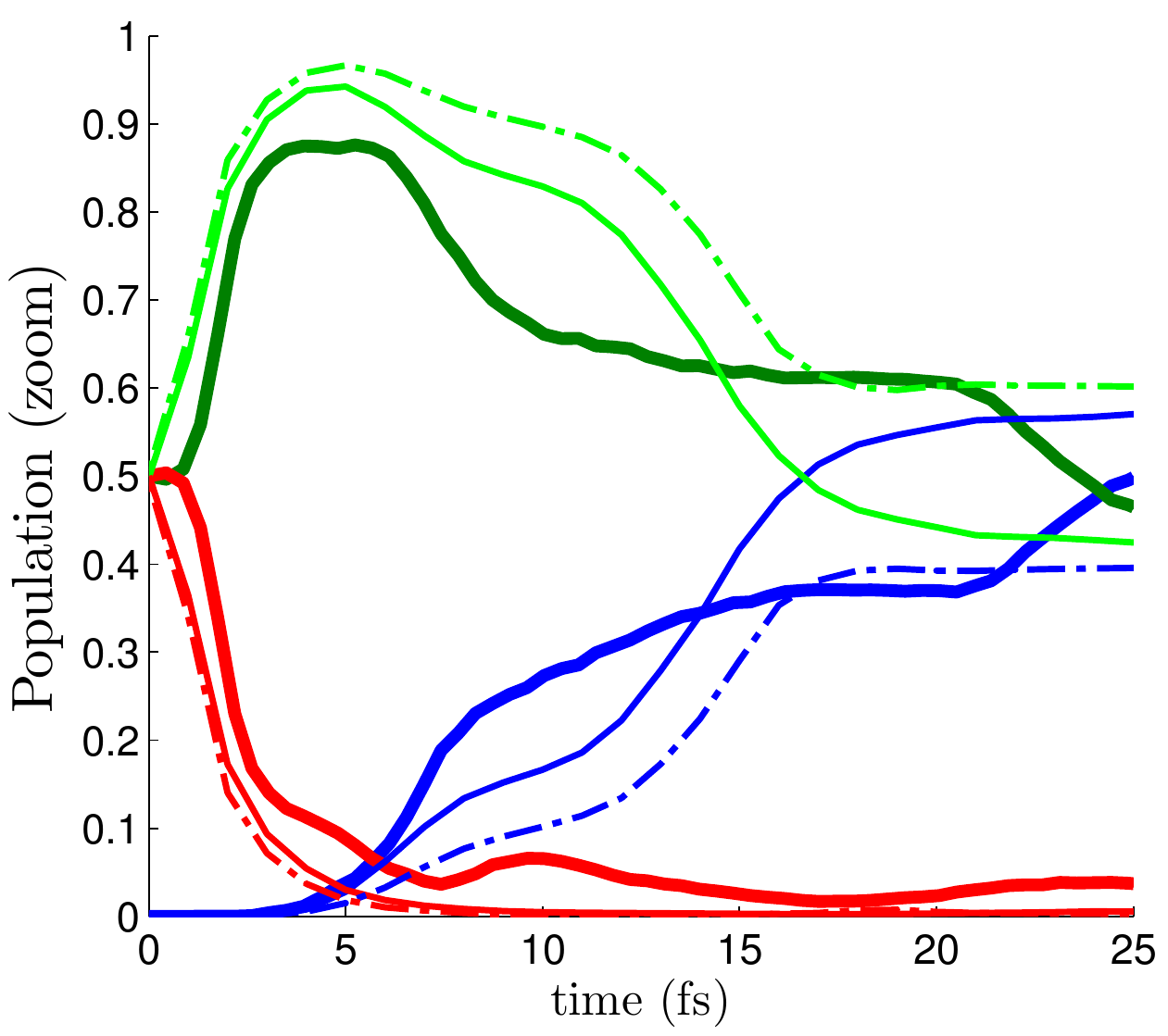}}
    \end{tabular}
    \caption{Populations of the three lowest adiabatic electronic levels (the ground state and the two lowest excited states) of NH$_3^+$ after ionization at $t=0$. The population of the first level is showed in blue, the population of the second second level in green and of the third level in red, respectively. The thick lines represent to the present results obtained with the LZ surface-hopping algorithm. The solid and the broken thin lines represent the rusults of the time dependent wave packet calculations from Refs.~\cite{Viel-etc:2006jcp, Viel-etc:2008cp}, respectively. The lower panel show the results of the first $25$ fs. The surface hopping results were obtained using $1000$ trajectories.}
    \label{fig:Pop}
\end{figure}

Figure \ref{fig:Pop} compares the adiabatic populations of the three lowest levels of NH$_3^+$ obtained with the probabilistic algorithm to the results of quantum calculations reported in Refs.~\cite{Viel-etc:2006jcp, Viel-etc:2008cp}. Up to a time of $100$ fs, the level populations obtained with the probabilistic hopping algorithm is roughly in between the two different quantum results. This confirms the good agreement between this algorithm and quantum solvers already observed in \cite{Belyaev-etc:2014jcp} for a simpler system.

The four different time scales of the nuclear dynamics are clearly seen in Fig.~\ref{fig:Pop}. 
The ultrafast nonadiabatic transitions from the level 3 (the second excited state) to the level 2 (the first excited state) occur on a 5 fs time scale due to the Jahn-Teller conical intersection. 
The main part of the fast transition from level 2 to level 1 (the ground state) takes place within a short time scale of 20 fs due to the pseudo-Jahn-Teller conical intersection. 
The relatively slow conversion of a part of a first-excited-state population to the ground state occurs within a time scale of 100 fs. 
After roughly 120 fs, the populations remain nearly constant for all three levels (see Fig. \ref{fig:Pop}). 
The remaining first-excited-state population is approximately $34$\%, the second-excited-state population roughly  $4$\%.
The first three time scales were found and explained in the quantum calculations \cite{Viel-etc:2006jcp, Viel-etc:2008cp} performed with different coordinates and based on the same PES\cite{Viel-etc:2006jcp}. 
The present results are in a good agreement with the quantum results. 

\begin{figure*}[h!]
    \begin{tabular}{l}                                                                                      
      \includegraphics[width=12cm,height=6cm]{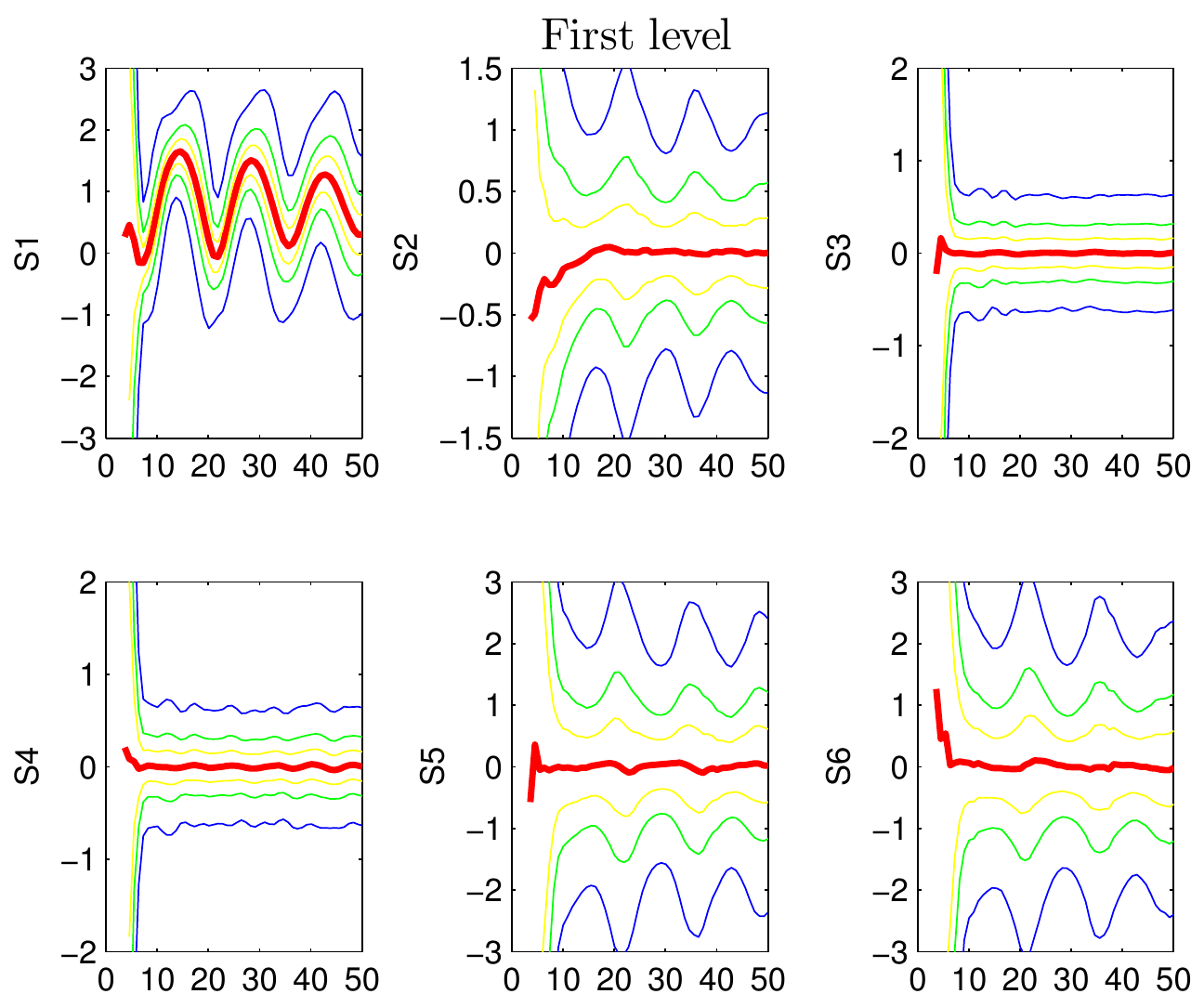}\\
      \includegraphics[width=12cm,height=6cm]{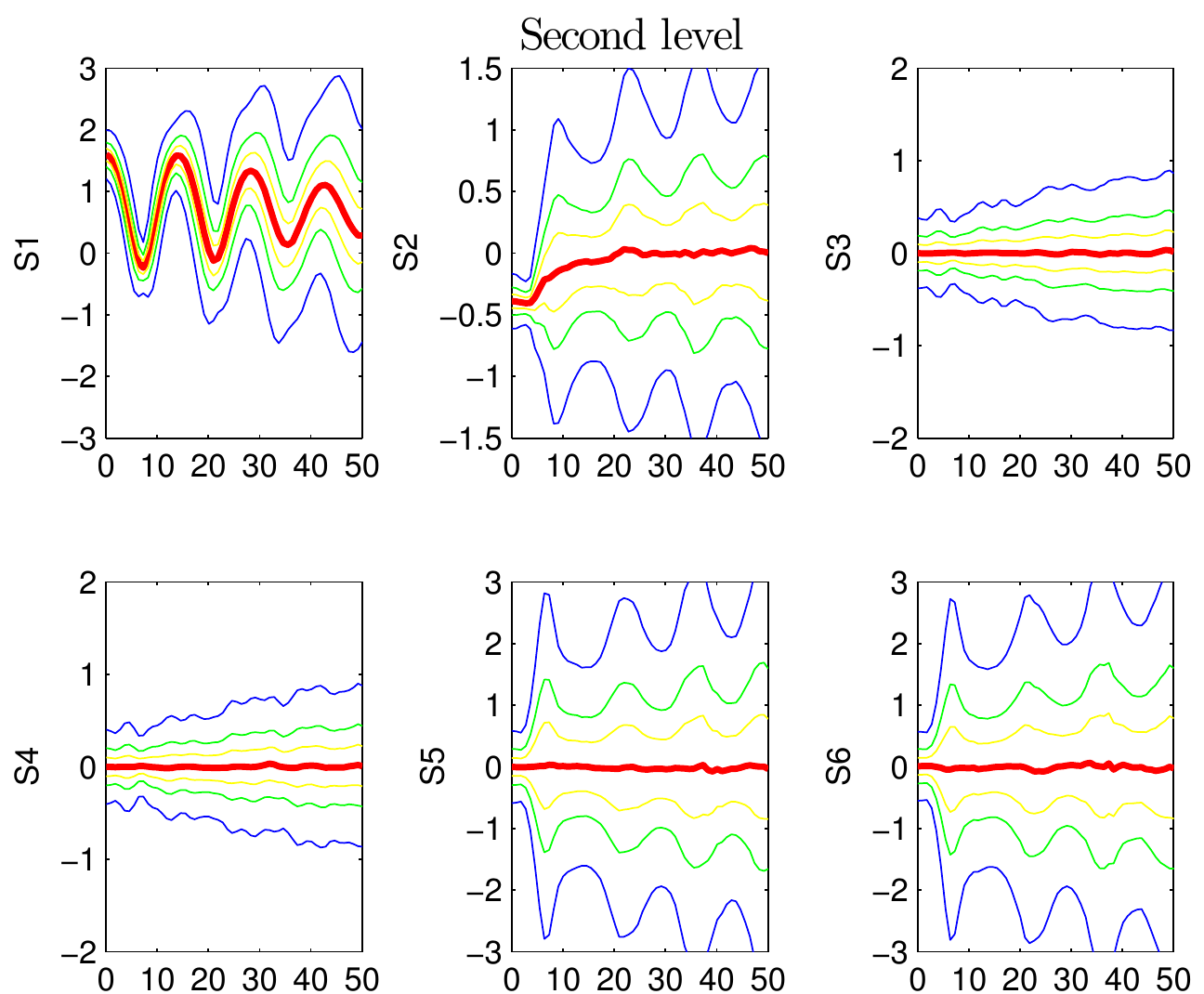}\\
      \includegraphics[width=12cm,height=6cm]{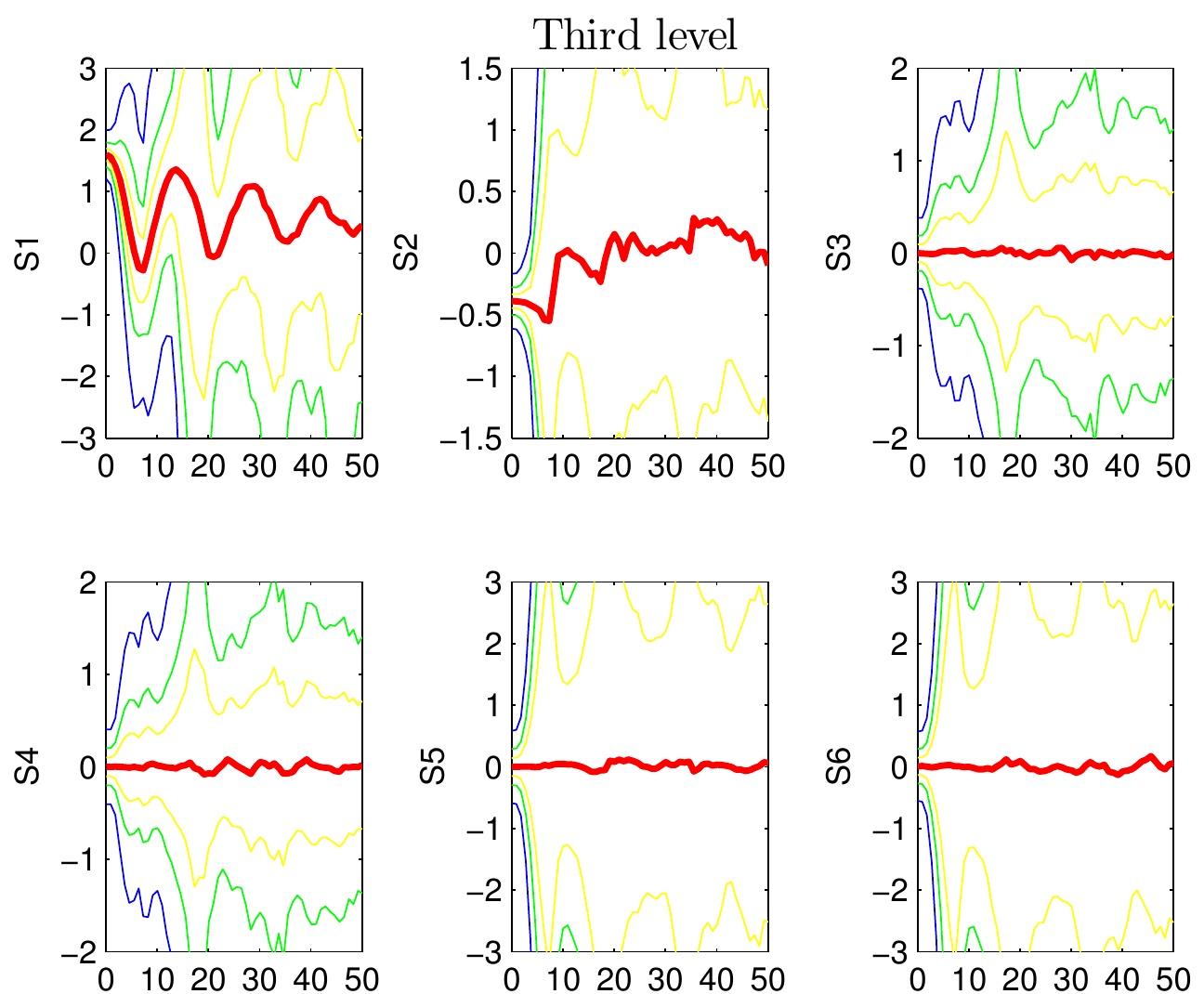}\\
     \end{tabular}
    \caption{Projection of the symmetry coordinates for each of the 3 levels as a function of time measured in femtoseconds. The panels on the first two rows refer to the first level, the 3rd and 4th row to the second level and the last two rows to the third level. The thicker red curve represents the average value of the coordinate while the thinner yellow, green and blue curves corresponds respectively to $0.5$, $1$ and $2$ standard deviation from the mean. It should be noted that the first level is not sufficiently populated during roughly the first $10$ fs leading to diverging values of the standard deviation from the mean.}
    \label{fig:SC}
\end{figure*}

The projection of the symmetry-adapted coordinates on the three different levels, see Fig.~\ref{fig:SC}, confirms what already observed in \cite{Viel-etc:2006jcp}. 
The analysis of the present results on the internal coordinate evolution shows that only within a short initial time interval the nuclear dynamics exhibits a vibronic motion followed by the spread of wave packets over wide ranges of a coordinate space.
In particular, the umbrella mode (coordinate $S_2$) takes roughly $20$ fs before spreading out reaching the mean value of zero corresponding planar configurations of the cation.

\subsection{Dynamical equilibrium}
\begin{figure}[h!]
      \resizebox{70mm}{!}{\includegraphics{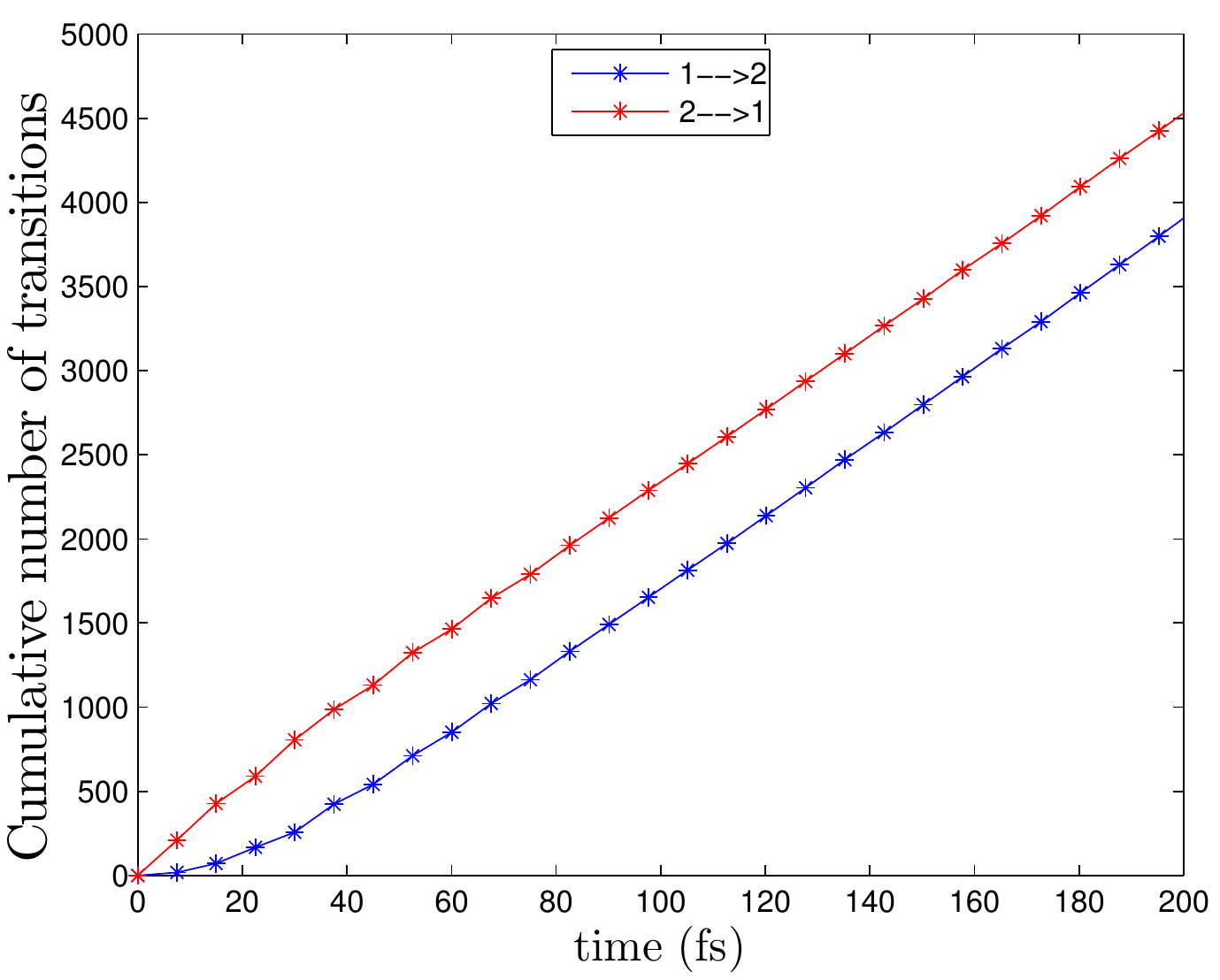}}
    \caption{Cumulative number of transitions from levels 1 and 2 (blue) and vice versa (red) versus time for the 1000 initial trajectory calculation. The number of trajectories hopping from the level 1 to 2 is roughly balanced by those one hopping from 2 to 1 as the two almost parallel lines indicate.}
    \label{fig:Trans}
\end{figure}

By looking at Fig.~\ref{fig:Pop}, it can be observed that the time interval going from  about $120$ up to $200$ fs is characterized by  dynamical equilibrium between the populations of the first, the second and the third levels. This is made more evident by plotting the cumulative number of transitions as a function of time between the first and second levels (see Fig. \ref{fig:Trans}).

In order to further characterize the dynamical equilibrium, the mean lifetime of each trajectory on each of the three levels was computed. In particular, each trajectory was partitioned in consecutive time intervals each representing the time spent by the trajectory on a given level. To compute the mean lifetime of a trajectory, say on level 1, we considered only  the time intervals specific to level 1 and computed the mean value.      

The mean lifetime was then averaged over all trajectories in the time interval  $t\in[60,200]$ fs, corresponding to the long time scale on which the dynamical equilibrium is established. The mean lifetimes reported in Table \ref{tab:AVLF} reveal that, on average, a trajectory spends about 3~fs on the third level, about $11$ fs on the second level and about $23$ fs on the first level. The short mean lifetime of a trajectory on the second and the third levels can provide a possible explanation why fluorescence was not detected in experiments performed on this system. 
Since fluorescence occurs on time scales of the order of nanoseconds, a time interval of 11 fs is too short to allow the detection of the emitted photons.

\begin{table}[h!]
\caption{Average and standard deviation of the time (in femtoseconds) spent by a trajectory on each of the three levels.}
\label{tab:AVLF}
\begin{tabular}{|c|c|c|}
\hline
\multicolumn{3}{|c|}{Average Lifetime (fs)}\\ 
\hline
 &mean & std\\ \hline
 First level & 22.61 & 0.68\\
Second level & 11.12 & 0.13\\
 Third level & 2.68 & 0.13\\
\hline
\end{tabular}
\end{table}

\subsection{Nonadiabatic regions}
\begin{figure*}
  \begin{center}
      \begin{tabular}{ll}
      \includegraphics[width=7cm,height=5.5cm]{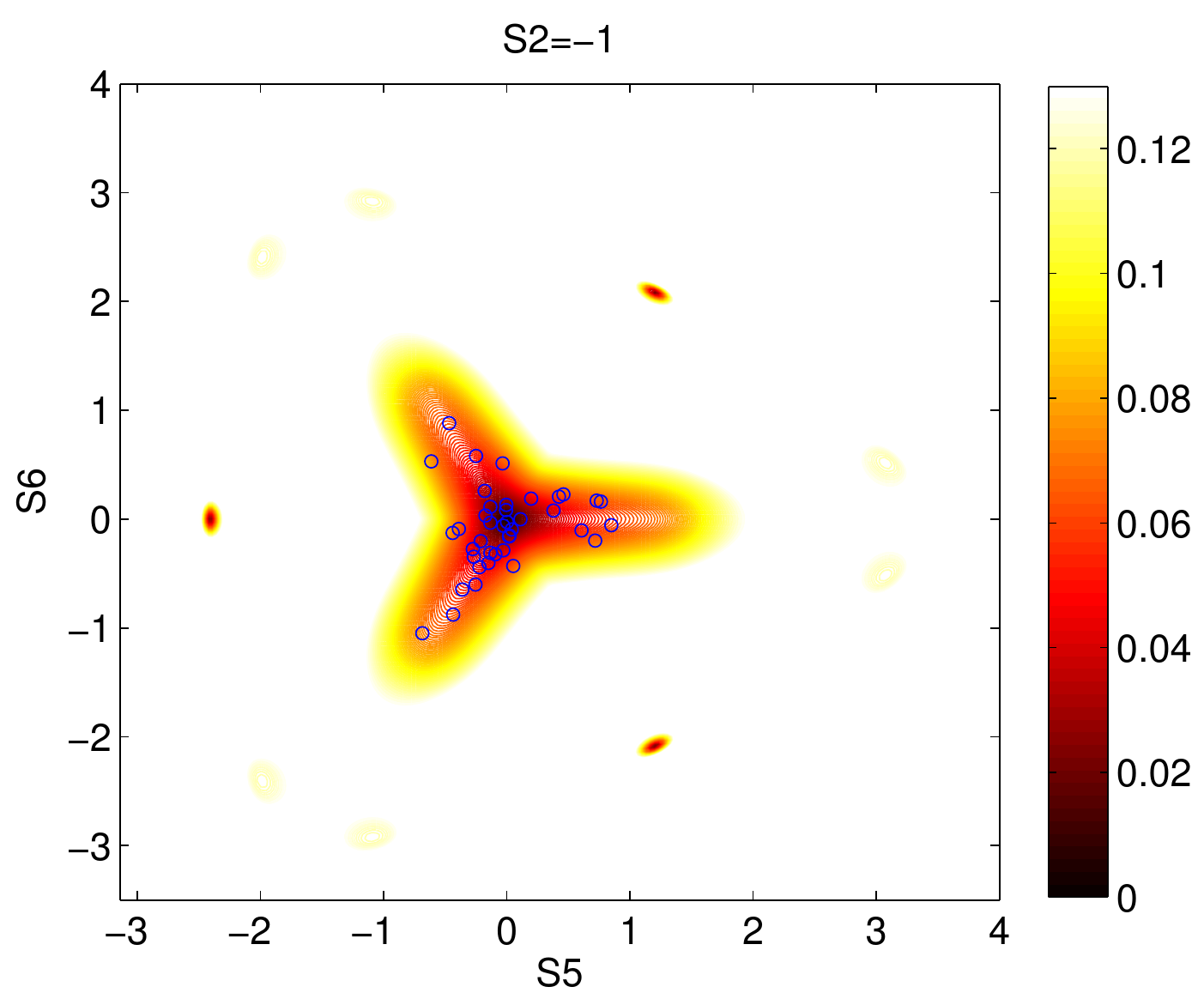}&
      \includegraphics[width=7cm,height=5.5cm]{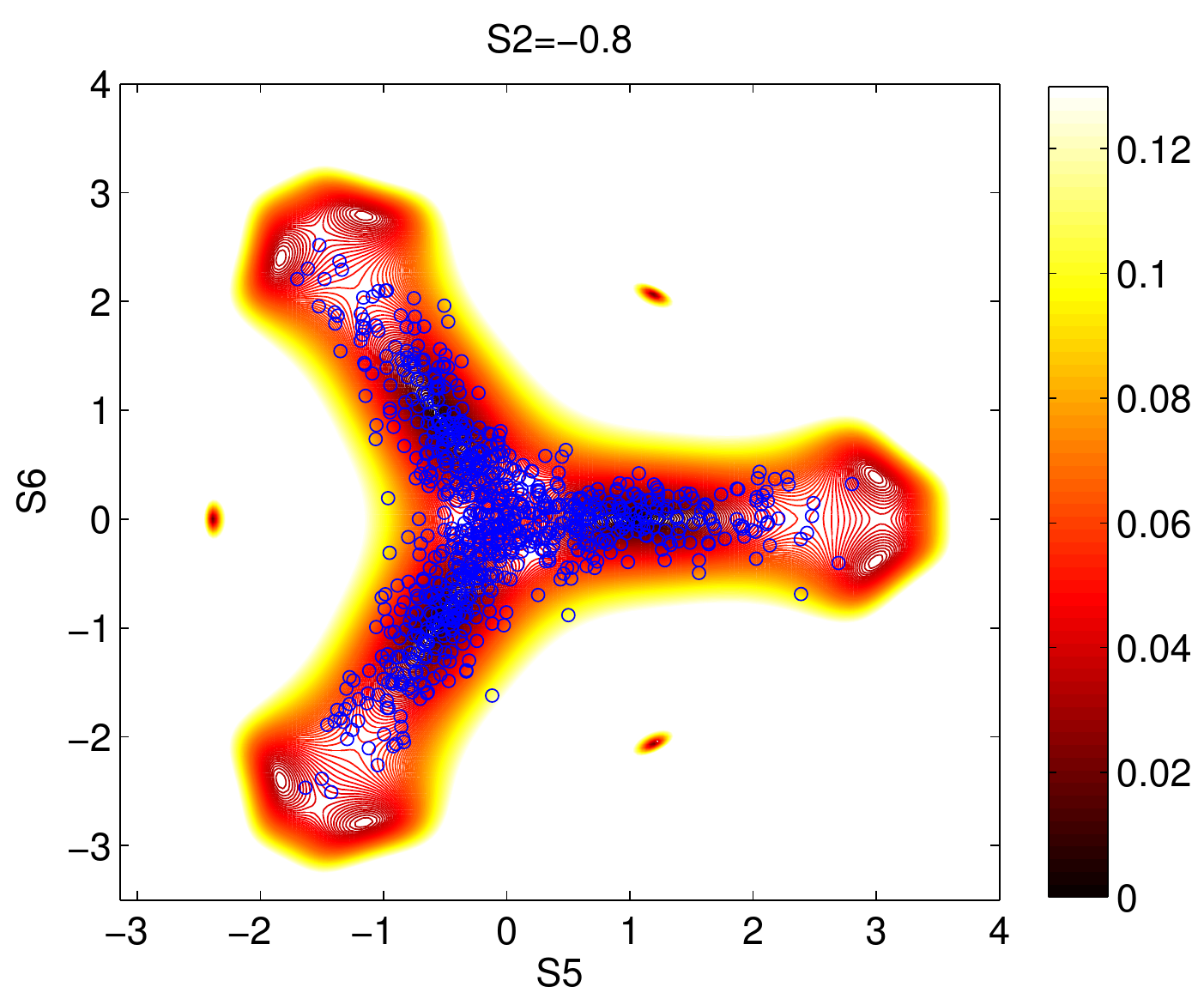}\\
      \includegraphics[width=7cm,height=5.5cm]{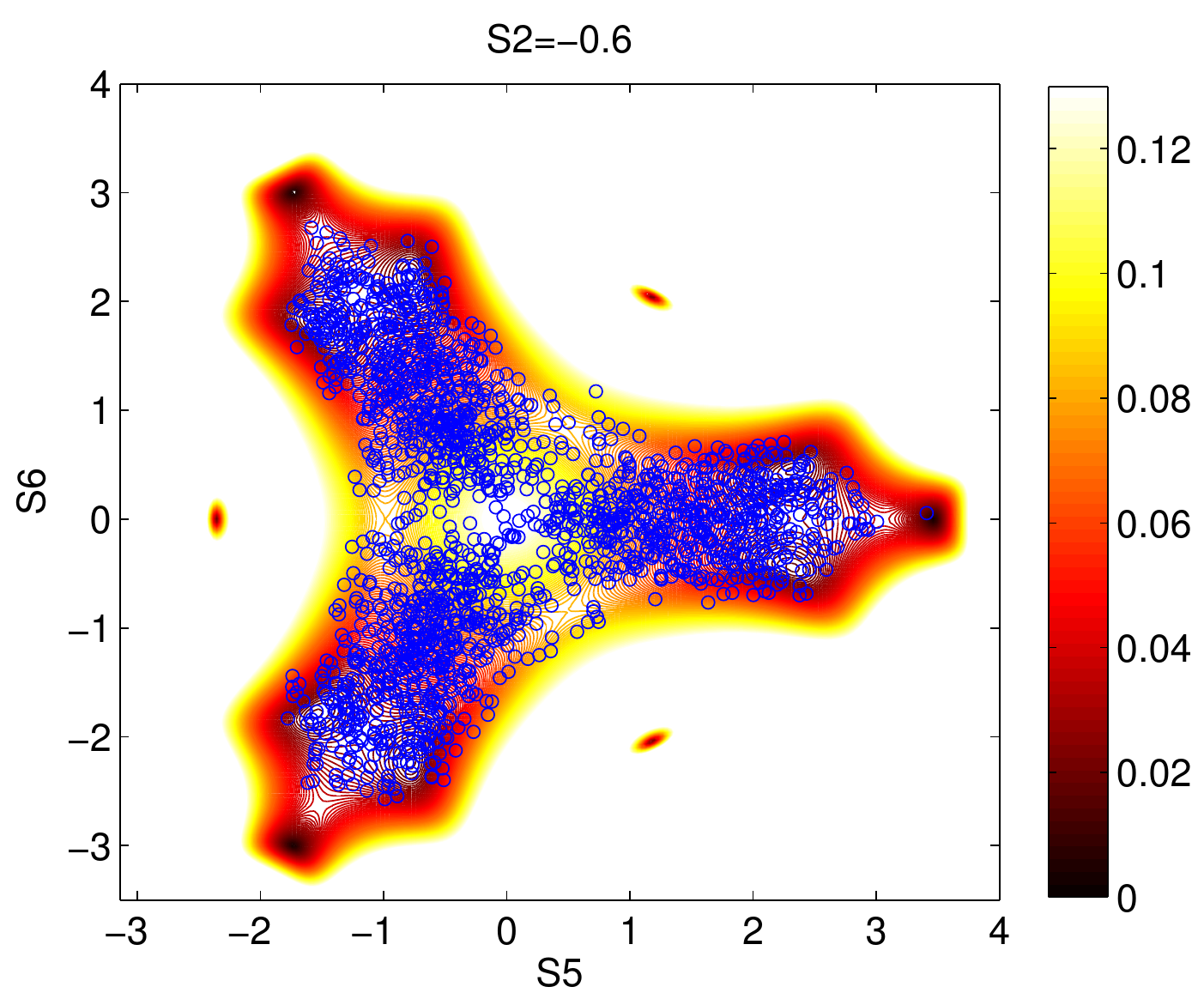}&
      \includegraphics[width=7cm,height=5.5cm]{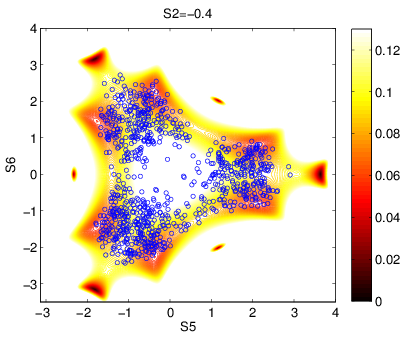}\\
      \includegraphics[width=7cm,height=5.5cm]{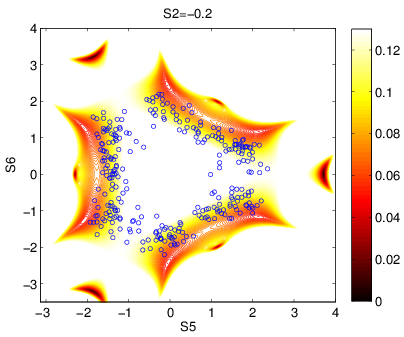}&
      \includegraphics[width=7cm,height=5.5cm]{TrPtsGapS2_6.png}\\	
     \end{tabular}
    \caption{\small{Gap between the first and second level as a function of $S_2$, $S_5$ and $S_6$ projected into the subspace $S_1=S_3=S_4=0$. The color code is such that dark red colors represent a small value of the gap. The blue dots are transition points that have been selected according to the value of $S2$ on intervals of length $0.2$ centered around the value reported on top of each panel. For instance, in the upper left panel are reported all points with $S_2 \in [-1.1, -0.9]$ and arbitrary value of the other five symmetry reduced coordinates.}}
    \label{fig:TrRegions}
  \end{center}
\end{figure*}

In this section the nonadiabatic regions involved in the transitions between the first and the second levels are described. In particular, the distribution of the hopping points in the six-dimensional space of the symmetry-adapted coordinates is analyzed. With the help of a technique for dimensionality reduction known as diffusion maps \cite{CoifmanLafon:2006}, it is possible to see that the nonadiabatic regions where transitions occur can be described with good approximation by considering only the coordinates $S_2$, $S_5$ and $S_6$.
In Fig. \ref{fig:TrRegions} the projection of the gap function between levels 1 and 2 in the subspace defined by $S_1=S_3=S_4=0$ is diplayed together with the marginal distribution of points into this subspace (blue dots). As can be noticed there is a good agreement between the conical intersections and the transition points, confirming the scarce relevance of $S_1, S_3, S_4$ in characterizing the hopping distribution points.

\begin{figure}[h!]
  \begin{center}
      \begin{tabular}{l}                                                                                      
      \resizebox{70mm}{!}{\includegraphics{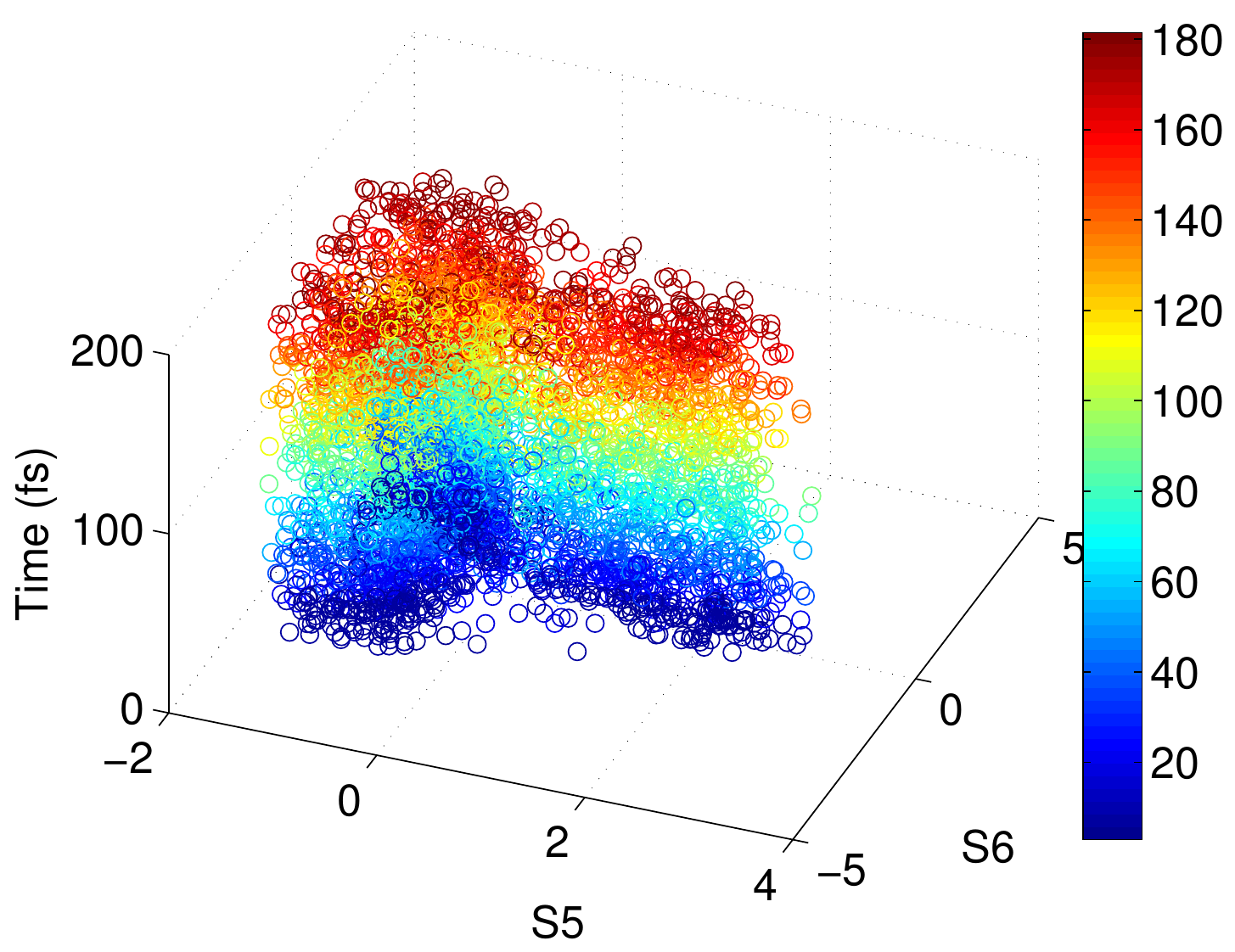}}\\
      \resizebox{70mm}{!}{\includegraphics{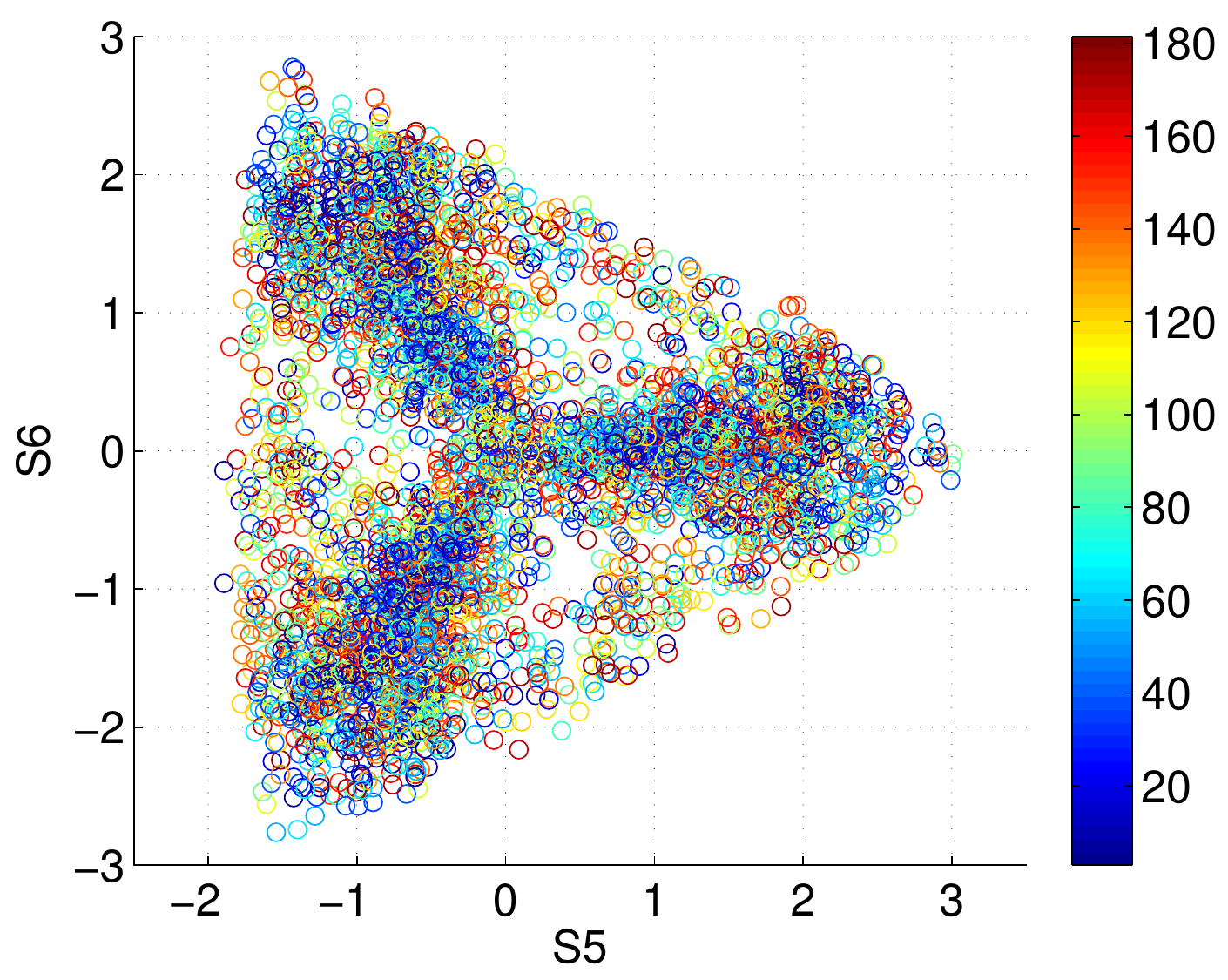}}
     \end{tabular}
    \caption{\small{Transition points as a function of time. In order to facilitate the visualization, the points are colored according to the time scale. As can be seen from the upper panel and the projection into the $S_5-S_6$ plane in the lower panel, the three transition regions described in the text are roughly active at the same time.}}
    \label{fig:shape}
  \end{center}
\end{figure}

Figure \ref{fig:TrRegions} shows that there are three active regions in the $S_5-S_6$ plane through which the dynamical equilibrium is taking place and that the shape of these regions is depending heavily on the value of the umbrella mode $S_2$.
It is natural to ask whether these three regions are relevant at different times during the simulation. 
By looking at figure \ref{fig:shape}, showing the transition points on the $S_5-S_6$ plane as a function of time, it is evident that the three nonadiabatic regions defined by figure \ref{fig:TrRegions} are roughly active at the same time. Particles passing through the nonadiabatic region at a given time are in fact characterized by different values of $S_2$, $S_5$ and $S_6$ so to reproduce a superposition of the panels in Fig. \ref{fig:TrRegions}.  

Although nonadiabatic transitions occur mainly in the vicinity of the conical intersections, only a few transitions take place exactly at the conical intersections. 
So, the nonadiabatic regions along trajectories are more of the avoided-crossing type. 
The present LZ surface-hopping approach handles both types of nonadiabatic transitions, the avoided-crossing and the conical-intersection ones.

\subsection{Consistency of the results}
The use of the probabilistic surface hopping algorithm described above requires averaging over different realizations of the dynamics obtained with identical initial condition in phase space. The populations of the levels in Fig. \ref{fig:Pop} and the number of transition in Fig. \ref{fig:Trans} are obtained by averaging the results of 10 different realizations. The maximum value over time of the standard deviation of the populations are $\sigma_{1}=0.0266,\sigma_{2}=0.0259$ and $\sigma_{3}=0.0097$, respectively. 

\begin{figure}[h!]
      \resizebox{70mm}{!}{\includegraphics{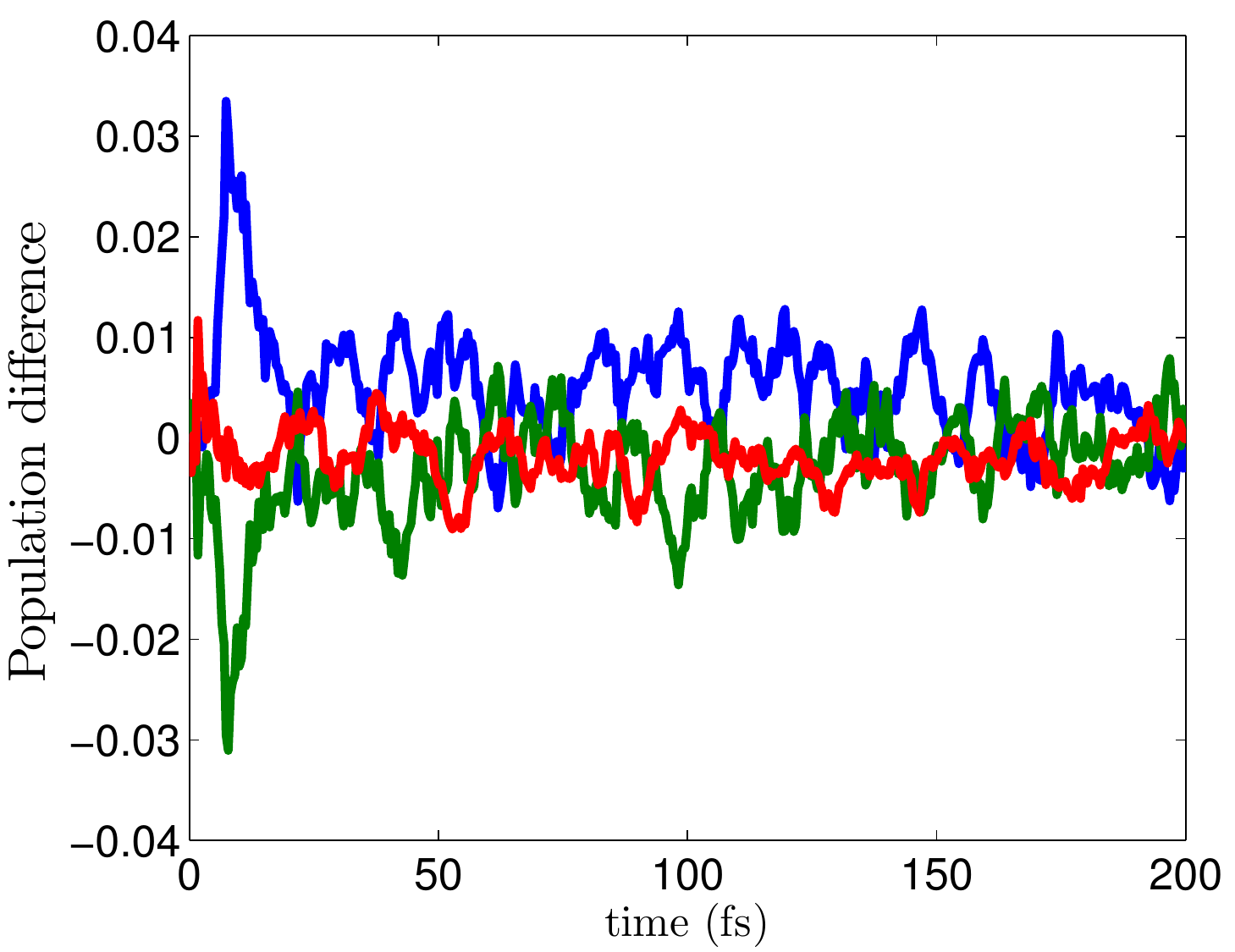}}
    \caption{Difference between the population of the levels obtained with 2000 and 1000 initial trajectories respectively. The first level is reported in blue, the second in green and the third excited level in red respectively.}
    \label{fig:Pop1000}
\end{figure}

In order to be sure that the Gaussian Wigner function relative to the initial wave packet was sampled with sufficient accuracy, we performed the calculation with 1000 and 2000 initial trajectories. The results for the population of the levels are reported in Fig. \ref{fig:Pop1000}. As can be observed, no substantial difference in between the two samples is noticed.
\section{Conclusions}
In the present work, the Landau-Zener type classical-trajectory surface-hopping algorithm \cite{Belyaev-etc:2014jcp} has been applied to the nonadiabatic nuclear dynamics of the ammonia cation after photoionization of the ammonia molecule. 
The algorithm employs an adiabatic-potential-based formula \cite{BelyaevLebedev:2011} for determination of LZ nonadiabatic transition probabilities along each classical trajectory. The algorithm only requires the information about adiabatic potential energy surfaces. 
For the ammonia cation, a six-dimensional three-sheeted PES is available \cite{Viel-etc:2006jcp}, which includes Jahn-Teller and pseudo-Jahn-Teller conical intersections between the ground and the two lowest excited states. 
This adiabatic multi-sheeted PES allowed us to study the nonadiabatic nuclear dynamics of the ammonia cation in detail for time scales which were not accessible for the quantum wave-packet calculations of Refs. \cite{Viel-etc:2006jcp, Viel-etc:2008cp}.
The time evolution of the populations of the ground and two lowest excited adiabatic electronic states after photoionization to the excited states was calculated up to 200 fs. 
The time dependence of the mean internal symmetry-adapted coordinates was calculated as well. 
The present classical surface hopping results for both the populations and the coordinates are in a good agreement with the quantum calculations \cite{Viel-etc:2006jcp, Viel-etc:2008cp} available for the first 100 fs time interval based on the same PES.

The present calculations reveal four different time scales in the nuclear dynamics for the PES of Ref.~\cite{Viel-etc:2006jcp}. 
The ultrafast nonadiabatic transitions from the second excited state to the first excited state occur on a 5 fs time scale due to the Jahn-Teller conical intersection. 
The main part of the fast transition from the first excited state to the ground state takes place within a short time scale of 20 fs due to the pseudo-Jahn-Teller conical intersection. 
The relatively slow conversion of a part of a first-excited-state population to the ground state occurs within a time scale of 100 fs. 
After roughly 120 fs, the populations remain nearly constant due to a dynamical equilibrium between all three states. 
For example, the remaining first-excited-state population is $\approx 34$\%.
The first three time scales are in agreement with the results of quantum calculations  \cite{Viel-etc:2006jcp, Viel-etc:2008cp}. 
Overall, the present results are in an excellent agreement with the quantum results.

The analysis of the internal coordinate evolution shows that only within a short initial time interval the nuclear dynamics exhibits a vibration motion followed by the spread of wave packets over wide ranges of a coordinate space.
The present calculations determine the locations of regions in nuclear coordinate space where nonadiabatic transitions occur. 
They are mainly in the vicinity of the conical intersections, although only a few transitions take place exactly at the conical intersections, so the nonadiabatic regions along trajectories are more of the avoided-crossing type. 

The classical trajectory calculations explain the experimental evidence that there is no photon emission from the excited state to the ground state of the ammonia cation. 
Although there are substantial remaining populations of the excited states, mainly of the first excited state, these populations represent a dynamical equilibrium due to nonadiabatic (non-radiative) transitions between all three states. 

Thus, the Landau-Zener type classical-trajectory surface-hopping algorithm is an efficient tool for studying nonadiabatic nuclear dynamics. It yields reliable results and provides physical insight into the dynamics of complex photophysical relaxation processes.


\appendix
\section{Symmetry adapted coordinates}
For completeness, we report below the definition of the symmetry adapted coordinates as appearing in \cite{Viel-etc:2006jcp, Viel-etc:2008cp}. If we indicate with $\overrightarrow{NH_{i}}$ the position of  $H_{i}$ in the system of reference of the $N$ atom and with  $h_{i}$ the unit vector $\overrightarrow{NH_{i}}/\parallel\overrightarrow{NH_{i}}\parallel$. Then the trisector $\hat{t}$ is the unit vector defined by the relation $\hat{t}\cdot h_{1}=\hat{t}\cdot h_{2}=\hat{t}\cdot h_{3}.$

The symmetry adapted coordinates are defined in terms of the following quantities: 
\begin{itemize}
 \item The three bond lengths $r_{1},r_{2},r_{3}$ between the $N$ atom and the three hydrogen atoms
 \item The angle $\beta$ in between the trisector $\hat{t}$ and any of the $h_{i}$
 \item The three projections, $\alpha_{1}, \alpha_{2}, \alpha_{3}$, of each HNH angle on the plane perpendicular to the trisector.
\end{itemize}

The symmetry adapted coordinates are defined in Table \ref{tab:SAC} as displacements with respect to the reference geometry characterized by $\beta=\pi/2$, $\alpha_{i}=2\pi/3$, and $\parallel\overrightarrow{NH_{i}}\parallel=1.02300190$~\AA  \ for every $i$. 
\begin{table}[h!]
\caption{Definition of the symmetry-adapted coordinates.}
{\renewcommand{\arraystretch}{2}
\begin{tabular}{|c|c c c|}
\hline
\multicolumn{4}{|c|}{\small{Symmetry Adapted Coordinates}}\\ 
\hline
\hline
Symmetric stretch &$S_{1}$ & = & $\frac{1}{\sqrt{3}}(\Delta r_{1}+ \Delta r_{2} +\Delta r_{3})$ \\
Umbrella mode &$S_{2}$ & = & $\Delta \beta$ \\
Asymmetric stretch &$S_{3}$ & = & $\frac{1}{\sqrt{6}}(2\Delta r_{1}- \Delta r_{2}- \Delta r_{3})$ \\
Asymmetric stretch &$S_{4}$ & = & $\frac{1}{\sqrt{2}}(\Delta r_{2}- \Delta r_{3})$ \\
Asymmetric bend &$S_{5}$ & = & $\frac{1}{\sqrt{6}}(2\Delta \alpha_{1} -\Delta \alpha_{2} -\Delta \alpha_{3})$ \\
Asymmetric bend &$S_{6}$ & = & $\frac{1}{\sqrt{2}}(\Delta \alpha_{2}-\Delta \alpha_{3})$ \\
 \hline
\end{tabular}}
\label{tab:SAC}
\end{table}

\section{Diffusion Maps}
Diffusion maps\cite{CoifmanLafon:2006} is a technique commonly used to map a given data set of points $\{x_{i}\}$ defined in $\R^{n}$ into a lower dimensional space $\R^{m}$ with $(m<n)$ in a way such that the geometric structure of the original dataset is efficiently represented in the new coordinates. 
The first step in determining a diffusion map is to define a local measure of similarity between points:
$$
K_{ij}=\exp\left(-\frac{|x_{i}-x_{j}|^{2}}{\alpha}\right) \, , 
$$
where $\alpha$ is a parameter specifying the lenght scale of the neighborhood to which $x_{i}$ and $x_{j}$ should belong. 
By normalizing to one each row of the diffusion kernel $K$, a Markov process with transition matrix 
$$
P=M^{-1}K
$$ 
is defined on the graph associated to the data set $\{x_{i}\}$), where $M_{ij}=\delta_{ij}\sum_{l}K_{il}$. 

\begin{figure}[h!]
  \begin{center}
      \resizebox{55mm}{!}{\includegraphics{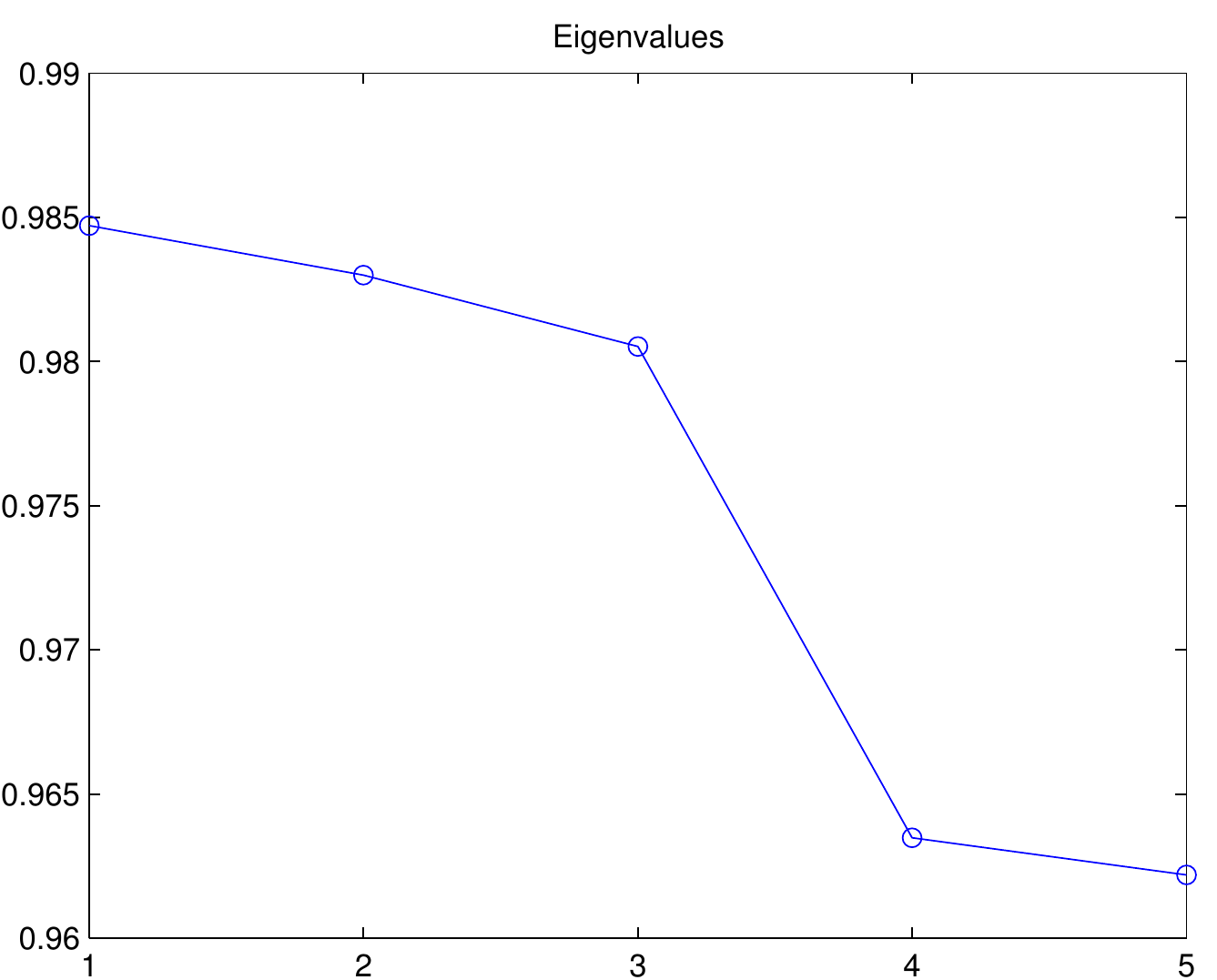}}
     \caption{\small{First five non-trivial eigenvalues of the matrix $P$.}}
    \label{fig:Eigenval}
  \end{center}
\end{figure}

Let $\mu_{1}\ge\ldots\ge \mu_{n}$ denote the eigenvalues of the transition matrix $P$. Since $P$ is a stochastic matrix, $\mu_{1}=1$ and all the remaining eigenvalues of $P$ have modulus smaller than one,
and the modulus of the eigenvalues of $P$ is connected to the number of relevant dimensions characterizing the set $\{x_{i}\}$.
Analyzing the spectrum of the matrix~$P$ associated with the hopping trajectories of the ammonia cation system, we notice a spectral gap between the third and the fourth non trivial eigenvalues (see Fig. \ref{fig:Eigenval}).

Due to the physical relevance of the six symmetry adapted coordinates $S_1,\ldots,S_6$ we assume that a subset of three of them could well reproduce the data set $\{x_{i}\}$ of hopping positions. 
Using the three eigenvectors associated to the three dominant eigenvalues ($\mu_{2},\mu_{3}$ and $\mu_{4}$),  we find that 
$S_{5}$, $S_{6}$ and $S_{2}$ are the three most relevant coordinates, while 
%
the distribution of transition points should not change much for small changes of $S_1$, $S_3$, $S_4$.

\begin{acknowledgements}
This research has been supported by the German Research Foundation (DFG) and the Russian Foundation for Basic Research (RFBR), grant for 
international collaboration \# LA 2316/3-1 (DFG) and \# 14-03-91337 (RFBR).
\end{acknowledgements}


\end{document}